\documentclass[a4paper,10pt]{paper}
\usepackage[english]{babel}
\usepackage{natbib}
\usepackage{amssymb}
\usepackage[centertags]{amsmath}
\usepackage{graphicx,tabularx}
\usepackage[latin1]{inputenc}
\usepackage{float}

 \newtheorem{lem}{Lemma}
 \def\mm#1{\ensuremath{\boldsymbol{#1}}}
\begin{document}

\title{Using Integrated Nested Laplace Approximation for Modeling Spatial Healthcare Utilization}
\author{Sauleau Erik A. \\
   \emph{Department of Biostatistics, University of Strasbourg, France} \\
   \emph{ea.sauleau@unistra.fr} \\
   Mameli Valentina \\ \emph{Department of Mathematics, University of Cagliari, Italy} \\
   Musio Monica \\ \emph{Department of Mathematics, University of Cagliari, Italy}}

\maketitle

\begin{abstract}

In recent years, spatial and spatio-temporal modeling have become an important area of research in many fields (epidemiology,
environmental studies, disease mapping). In this work we propose different spatial models to study  hospital recruitment,
including  some potentially explicative variables. Interest is on the distribution per geographical unit of the ratio between the
number of patients living in this geographical unit $i$, say $y_i$, and the population, $N_i$ in the same unit.
Models considered  are within the framework of Bayesian Latent Gaussian models \citep{Fahrmeir2001a}. Our response variable $y_i$
is assumed to follow a binomial distribution, with logit link,  whose parameters are the population $N_i$ in the
geographical unit $i$ and the corresponding relative risk  $\pi_i$. The structured additive predictor $\eta_i$ accounts for
effects of various covariates in an additive way: $\eta_i = \alpha +
\sum_{j=1}^{n_f}{f^{(j)}(\mm{u}_{ji})}+\sum_{k=1}^{n_{\beta}}{\beta_k z_{ki}} + \epsilon_i$. Here, the $f^{(j)}(\cdot)$s are
unknown functions of the covariates $\mm{u}$ (which also includes a spatial effect)  the $\beta_k$s represent the linear effect
of covariates $\mm{z}$ and the $\epsilon_i$s are unstructured terms.  To approximate posterior marginals, which not available in
closed form, we use integrated nested Laplace approximations (INLA) \citep{Rue2009}, recently proposed for approximate Bayesian
inference in latent Gaussian models. INLA has the advantage of giving very accurate approximations and being faster than McMC
methods when the number of parameters does not exceed  6 (as it is in our case). Model comparisons are assessed using DIC
criterion \citep{Spiegelhalter2002}.
\end{abstract}

\section{Introduction}

Analysis of spatial hospital utilization patterns is a fundamental requirement for effective health services planning and hospital management.
Two different approaches are used in these studies:
\begin{enumerate}
 \item[(a)] The descriptive approach consists essentially in recruitment mapping. Maps can be produced for different population groups,  for example distinguishing males and females or age categories. The criteria for these distinctions are clinical and not statistical-based. Moreover, the maps need a spatial smoothing in order to be more easily interpretable.
 \item[(b)] The explicative approach, the aim of which is to find what variable(s) can explain differences in the observed recruitment. Different statistical models are constructed and compared in order to highlight these variables. Furthermore, models can be used for simulating the effect on the recruitment of the modification of some of these variables (for example what happens if a new road decreases the access time to a certain hospital from some cities of a region?). Models can also be used to take into account population structure evolution on the predictions of the recruitment.
\end{enumerate}

The study of the recruitment, for instance, for a particular hospital, a given disease, or a particular age range, implies having a geographic reference of patient residence. Identifying the place of residence of patients may allow geocoding and subsequent use of geostatistical models for point processes (see \citet{Cressie1993}). The problem arises of determining the population at risk to be matched to each patient or patient group. Furthermore, the reliability of the exact address is not assured. Models for grouped data applied to geographic units are then meaningless. The address of each patient is reported to a geographical unit, in which a population at risk can be determined. Different possibilities for defining geographical units can be explored. The most detailed level available in years is the municipality ("commune", defined by the French National Institute for Statistics and Economic Studies, INSEE). The French National Geographic Institute (IGN) calculates the "ce
 ntroid" of each municipality, that is to say virtual centers taking into account the shape of the municipality. These centroids can also be used to locate recruited cases. The municipality of residence is a dataset item that is always present in the hospital information systems, probably even updated if necessary at each patient visit. In the rest of this paper we will retain the municipal level but generally  we will speak of the geographical unit.

Concerning the hospital recruitment, the interest is on the distribution per geographical unit of the ratio between the number of
patients living in this geographical unit $i$, say $y_i$, and the population, number of persons "at risk" to visit an healthcare provider, $N_i$
in the same unit. We call this ratio, $\frac{y_i}{N_i}$, the standardized recruitment ratio (\textbf{SRR}). We assume that the response variable $y_i$
independently follows a binomial distribution whose parameters are the population $N_i$ and a particular risk per
unit $\pi_i$.

If $\textrm{logit}(\pi_i) = \eta_i$, we then have $\pi_i=\frac{e^{\eta_i}}{1+e^{\eta_i}}$. The covariates  enter the model additively through the predictor $\eta_i$,
$$\textrm{logit}(\pi_i) = \eta_i =\mu + \sum_{j=1}^{n_f}{f^{(j)}(\mm{u}_{ji})} + \sum_{k=1}^{n_{\beta}}{\beta_k z_{ki}} + \mm{\epsilon}_i. $$
Here, the $f^{(j)}(\cdot)$s are unknown functions of $n_f$ covariates in $\mm{u}$ (including also a spatial effect), the $\beta_k$s represent the linear effect of $n_{\beta}$ covariates in $\mm{z}$ and the $\epsilon_i$s are unstructured terms. The model adopted is a structured additive regression model (StAR model), see \cite{Fahrmeir2001a}. In this model, the response variable $y_i$ is assumed to belong to an exponential family, where the mean $\mu_i$ is linked to a structured additive predictor $\eta_i$ through a link function $g(\cdot)$, so that $g(\mu_i)=\eta_i$. The structured additive predictor $\eta_i$ accounts for effects of various covariates in an additive way. This class of models can be complex and hierarchical, involving fixed and random effects and are particularly suited to Bayesian inference \citep{Gelman1995, Banerjee2004}, although in this context, the term "fixed" no longer has the classical meaning it has under purely frequentist inference. Our aim is to use the model above to explain spatial recruitment in Haute Alsace, a region in the north-east of France, using data from  the public hospital of Mulhouse, the biggest town of the region. Several alternative explicative variables  are considered. As is often found in disease mapping literature \citep{Bernardinelli1995, Rue2005} we have chosen Gaussian priors for $\mu$, $f(\cdot)$, $\beta$ and $\epsilon$. Our model is then a latent Gaussian model  and parameters  $\mm{x}=(\mu, f(\cdot), \beta, \epsilon)$ are called latent Gaussian variables. Hyperparameters $\mm{\theta}$ involved in prior elicitations are not necessarily Gaussian.  The common approach to inference for latent Gaussian models is Markov chain Monte Carlo (McMC) sampling. It is well known, however, that McMC methods tend to exhibit poor performance when applied to such models. Various factors explain this. First, the components of the latent field $\mm{x}$ are strongly dependent on each other. Second, $\mm{\theta}$ and $\mm{x}$ are also strongly dependent, especially when $n$ is large. Despite developments (see for instance \citet{Banerjee2008, Held2010}) for overcoming this poor performance, McMC sampling remains painfully slow from the end user's point of view.  To approximate posterior marginals,  we use integrated nested Laplace approximations (INLA) \citep{Rue2009,Rue2007}, recently proposed for approximate Bayesian inference in latent Gaussian models. INLA has the advantage of giving very accurate approximations and being faster than McMC methods when the number of parameters does not exceed  6 (as it is in our case). Model comparison and selection will be assessed using Deviance information criterion (DIC), see \citet{Spiegelhalter2002}. Implementation of space and space-time models with INLA are presented and explained in detail in \citet{Schroedler-Held2009,Schroedler-Held2009bis}.

The structure of the paper is as follows: in Section 2 we describe the data in detail. In Section 3 we introduce and justify the model used including details on assumptions on the priors and on the INLA method used for inference. Results obtained and comparisons of different models are shown in Section 4. We finish with a discussion in Section 5.

\section{Data description and explanatory analysis}

\begin{figure}[H]
 \centering
 \makebox{\includegraphics[scale=0.6]{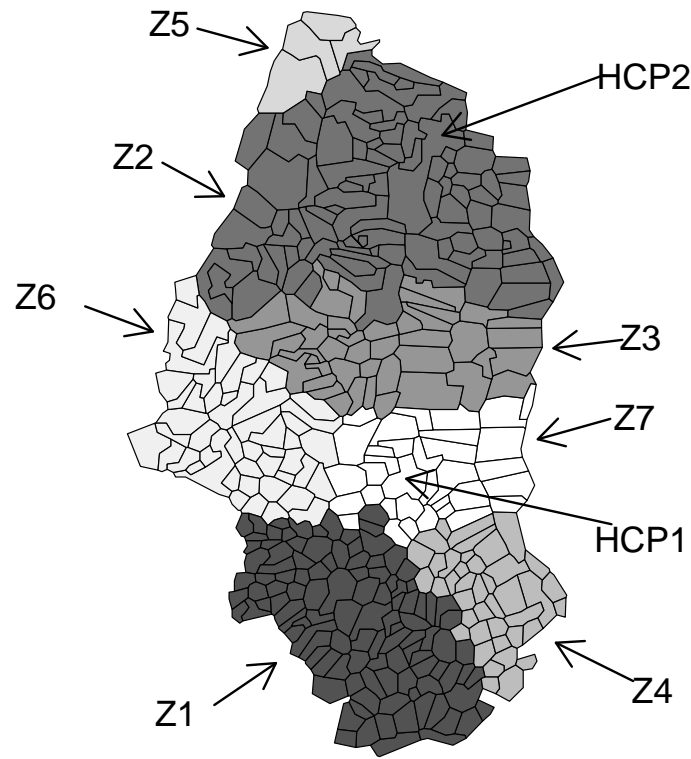}}
 \caption{\textbf{Proximity zones of the region (Z1: Altkirch, Z2: Colmar, Z3: Guebwiller, Z4: Saint-Louis, Z5: Sélestat, Z6: Thann, Z7: Mulhouse). The studied healthcare provider is HCP1 and the second provider is HCP2}}
 \label{fig:Zp68}
\end{figure}

Data are from the public hospital of Mulhouse (its location is shown on Figure (\ref{fig:Zp68})), the biggest town of the Haute Alsace region in north-east of France. This region, adjacent to Germany and Switzerland, is 3,525 $\textrm{km}^2$ and has 756,974 inhabitants (01/01/2010) in a very dense irregular lattice of 377 municipalities ("communes") which are the geographical units we use. The largest distance between the centroids of two geographical units is about 95 km.

\begin{figure}[H]
 \centering
 \makebox{\includegraphics[scale=0.6]{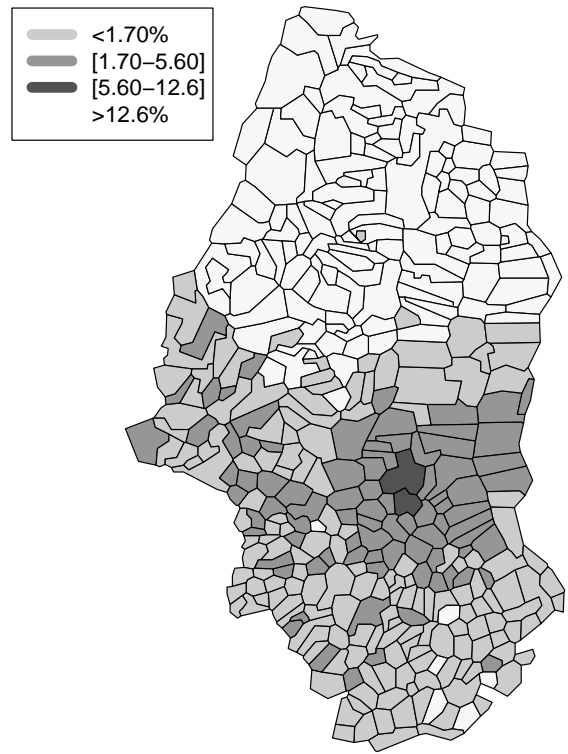}}
 \caption{\textbf{Observed recruitment ratio per geographical unit}}
 \label{fig:SRRObs}
\end{figure}

In 2008, all the 12 healthcare providers in the region recorded 182,487 visits (in- and outpatients). The hospital of Mulhouse recorded 48,747 among these (27\%). In this paper, we only consider the 33,682 inpatients. The distribution of the number of cases across the 377 geographical units is very heterogeneous: there are between 0 and 12,330 cases per geographical unit, the mean is 89 cases but the median is 17 with 99\% of the geographical units having less than 1,019 inpatients. Only 4 geographical units had more than 1,000 inpatients. Figure (\ref{fig:SRRObs}) represents the observed recruitment ratio per geographical unit (calculated as the number of inpatients which in a given geographical unit divided by the population of this geographical unit).

We have different potential explicative variables affecting recruitment. More precisely we have to deal with the following requirements:
\begin{enumerate}
 \item[(a)] Practitioners send their patients preferentially (except for some particular pathologies) to a given healthcare provider. These practitioners filter patients on several closed geographical units. This  means that the recruitment in a given geographical unit is more "similar" to that in a closed unit than that in another random unit in the region. This is the definition of spatial autocorrelation and we can assume that the use of statistical models taking into account the autocorrelation greatly improves the explanation of the recruitment.
 \item[(b)] The distance or the access time between the healthcare provider and the geographical unit of residence reflects the ease of access to this healthcare provider. The access time may have a greater influence on recruitment in the context of emergency. Specifically, interest is focused on measuring the attenuation of recruitment with distance. Random walks can be used to smooth this attenuation.
 \item[(c)] A recent French healthcare policy introduced the notion of "proximity zones". The region is divided into several of these zones, shown on Figure (\ref{fig:Zp68}), each centered by a healthcare provider to which its patients are recruited. But there are different levels of providers according to their technical capacities and competencies. A bigger provider also has to recruit patients (for various specific pathologies) into several of these subregions.
 \item[(d)] Some other covariates can also influence the recruitment, such as age, geographical characteristics or economic status of the geographical units, etc. Most of these covariates are beyond the topic of this paper. Herein, we test only two:
      \begin{enumerate}
       \item[(i)] The distance between each geographical unit and a second important healthcare provider (HCP2 on Figure (\ref{fig:Zp68})) assuming that patients living nearer this second provider will prefer to go there rather than the first.
       \item[(ii)] The density of practitioners in each geographical unit (for 1,000 inhabitants), assuming that a higher density will result in a higher recruitment.
      \end{enumerate}
  \end{enumerate}

This consideration lead us to consider Bayesian structured additive regression models (StAR models) \citep{Fahrmeir2001a}. We will present the adopted model in detail in the following section.

\section{Statistical Model}

We assume that the response variable  $y_i$, the number of observed cases in the $i$th geographical unit ($i=1,\cdot,377$) follows
a binomial distribution with parameters, $N_i$ and $\pi_i$, where $N_i$ indicates the population and  $\pi_i$ is the relative risk. Thus $y_i \sim
\textrm{Bin}(N_i, \pi_i)$. We consider the logit link and the following additive structure for the linear predictor:
 \begin{equation}\label{eq:eta} \textrm{logit}(\pi_{i}) = \eta_{i} = \mu+\sum_{a=1}^{n_f}{f^{(a)}(\mm{u}_{ai})} + \sum_{k=1}^{n_{\beta}}{\beta_kz_{ki}} + f^{(s)}_{i} + f^{(u)}_{i}. \end{equation}
\noindent Here, the $f^{(a)}(\cdot)$s are unknown functions of the covariates $\mm{u}$, the $\beta_k$s represents the linear
effect of covariates $\mm{z}$, $\textbf{f}^{(s)}$ is a spatially structured component and $\textbf{f}^{(u)}$ is a spatially
unstructured component. The unstructured spatial component can be used as a proxy for important environmental covariates not included in the
analysis.

We assume the following prior distributions:
\begin{itemize}
\item $\textbf{f}^{(a)}$ follows an intrinsic second-order random walk model with precision $\tau^{(a)}$,
$$\pi(\textbf{f}^{(a)}|\tau^{(a)})\propto(\tau^{(a)})^{\frac{n_f-2}{2}}\exp\left\lbrace -\frac{\tau^{(a)}}{2}\sum_{j=3}^{n_f}\left( f_{j}^{(a)}-2f_{j-1}^{(a)}+f_{j-2}^{(a)}\right)^{2} \right\rbrace.$$ In addition, on $f_{1}^{(a)}$ and $f_{2}^{(a)}$ are specified vague priors (for example uniform).
\item The model for the spatial structured component $\textbf{f}^{(s)}$ is an intrinsic conditional autoregressive process (or Markov Gaussian random field)~\citep{Besag1991a,Mollie1996}, \textbf{ICAR}, which assumes that, conditionally on the spatial effect across adjacent geographic units, the effect in a unit follows a normal distribution. The average of this distribution is the average of spatial effects in the surrounding units and its variance is proportional to the number of neighbors of this unit. If $f_i$ is the effect in the unit $i$ and $\textbf{f}_{-i}$ the effects in units other than $i$ of the study area, then the ICAR can be written:
$$f_i^{(s)}|\textbf{f}^{(s)}_{-i},\tau^{(s)}\sim \mathcal{N}\left( \frac{1}{n_{i}}\sum_{j\in\partial_{i}}f_{j}^{(s)},\frac{1}{n_{i}\tau^{(s)}}\right).$$
In this formula, $n_i$ is the number of units adjacent to each $i$ and $\partial_i$ represents the set of all of these adjacent units. The adjacency between units is most often  defined according to the
notion of common boundary i.e. are considered  as adjacent if two units share a common
border. The only parameter to estimate is then $\tau^{(s)}$, the precision parameter of the ICAR. In this model, one may ask whether any spatial effect of the data is taken into account by the ICAR. We thus can seek to distribute the residuals on each of the geographical units. The association of this residual "heterogeneity" and the autocorrelation is the model traditionally used in disease mapping risks and called "convolution prior" involving an intrinsic conditional autoregressive process (for autocorrelation) and a normal distribution by geographical unit (for heterogeneity). Then $\textbf{f}^{(u)}$ are independent zero-mean Gaussian with precision $\tau^{(u)}$~\citep{Besag1991a,Mollie1996}.
\end{itemize}
We will assign independent $\Gamma(0.001, 0.001)$ priors to the hyperparameters $(\tau^{(a)},\tau^{(s)},\tau^{(u)})^{T}$ and a $\mathcal{N}(0,0.01)$ prior to $\mu$ and to $\beta_k$. Latent Gaussian models are a subset of Bayesian additive models with a structured additive predictor, in which Gaussian priors are assigned to $\mu$, all $f(\cdot)$, $\beta$. Let $\mm{x}$ be the vector of all the $n$ Gaussian variables $\mu$, $f(\cdot)$ and $\beta$ and $\boldsymbol{\theta}$ the vector of hyperparameters, which are not necessarily Gaussian.  The main goal of a Bayesian inference method is to estimate the posterior distribution
\begin{equation}\label{Bayes}
\pi(x_{i}|\textbf{y})=\int \pi(x_{i}|\boldsymbol{\theta},\textbf{y})\pi(\boldsymbol{\theta}|\textbf{y})d\boldsymbol{\theta}.
\end{equation}
We present the INLA approach for approximating the posterior marginals of $\pi(x_{i}|\textbf{y})$, $i = 1,\cdots, n.$ The approximation is computed in three steps. The first step approximates the posterior marginal of $\boldsymbol{\theta}$ by using the Laplace approximation. The second step computes the Laplace approximation or the simplified Laplace approximation of $\pi(x_{i}|\textbf{y},\boldsymbol{\theta})$ for selected values of $\boldsymbol{\theta}$. The third step combines the previous two steps and uses numerical integration for retrieving the final estimate of equation (\ref{Bayes}).

\begin{description}
\item [First step]
The marginal posterior density $\pi(\boldsymbol{\theta}|\textbf{y})$ of the hyperparameters $\boldsymbol{\theta}$  in (\ref{Bayes}), is approximated in the following way:
\begin{equation*}
\tilde{\pi}(\boldsymbol{\theta} |\textbf{y})\propto \frac{\pi(\textbf{x},\boldsymbol{\theta} |\textbf{y})}{\tilde{\pi}_{G}(\textbf{x}|\boldsymbol{\theta},\textbf{y})}\Big{|}_{\textbf{x}=\textbf{x}^{*}(\boldsymbol{\theta})}
\end{equation*}
where  $\tilde{\pi}_{G}(\textbf{x}|\boldsymbol{\theta} ,\textbf{y})$ is the Gaussian approximation of  $\pi(\textbf{x}|\boldsymbol{\theta} ,\textbf{y})$ and  $\textbf{x}^{*}(\boldsymbol{\theta})$
is the mode of  $\pi(\textbf{x}|\boldsymbol{\theta} ,\textbf{y}).$\\
The main use of $\tilde{\pi}(\boldsymbol{\theta}|\textbf{y})$ is to integrate out the uncertainty with respect to $\boldsymbol{\theta}$ when approximating the posterior marginal of $x_{i}$. For this task it is sufficient to be able to select good evaluation points for the numerical
integration. We locate the mode of $\tilde{\pi}(\boldsymbol{\theta}|\textbf{y})$, by optimizing $\log{(\tilde{\pi}(\boldsymbol{\theta}|\textbf{y}))}$ with respect to $\boldsymbol{\theta}$. This can be done by using some quasi-Newton method.
Let $\boldsymbol{\theta}^{*}$ be the modal configuration, at $\boldsymbol{\theta}^{*}$ we compute the negative Hessian matrix $H$, using finite differences. Let $\Sigma =H^{-1}$, which would be the covariance matrix for $\boldsymbol{\theta}$ if the density were Gaussian. To help the exploration, we use standardized variables $\textbf{z}$ instead of $\boldsymbol{\theta}$. Let $\Sigma=V\Lambda V^{T}$ be the eigendecomposition of $\Sigma$, and define $\boldsymbol{\theta}$ via $\textbf{z}$:
$$\boldsymbol{\theta}(\textbf{z})=\boldsymbol{\theta}^{*}+V\Lambda^{\frac{1}{2}}\textbf{z}.$$ We explore $\log{(\tilde{\pi}(\boldsymbol{\theta}|\textbf{y}))}$ by using \textbf{z}-parameterization. We start from the mode $(\textbf{z}=0)$ and go in the positive direction of $z_{1}$ with step length $\delta_{z}$, say $\delta_{z}=1$, as long as
\begin{equation}
\log{(\tilde{\pi}(\boldsymbol{\theta}(0)|\textbf{y}))}-\log{(\tilde{\pi}(\boldsymbol{\theta}(z)|\textbf{y}))}<\delta_{\pi},\label{ric}
\end{equation}
where for example, $\delta_{\pi}=2.5$. Then we switch direction and do similarly. The other coordinates are treated in the same way.
Posterior marginals for $\theta_{j}$ can be obtained from $\tilde{\pi}(\boldsymbol{\theta}|\textbf{y})$ by using numerical integration but this is computationally demanding. Then we use the points that satisfy the equation (\ref{ric}) to construct an interpolant to $\log{(\tilde{\pi}(\boldsymbol{\theta}|\textbf{y}))}$ and compute marginals by using numerical integration from this interpolant.
\item [Second step]
We have now a set of weighted points $\left\lbrace \theta_{k}\right\rbrace $ and have to find an accurate approximation for the posterior marginal for the $x_{i}$s, conditioned these selected values of $\boldsymbol{\theta}$.
The density $\pi(x_{i}|\boldsymbol{\theta} ,\textbf{y})$ is approximated using the Laplace approximation defined by:
\begin{equation}
\tilde{\pi}_{LA}(x_{i}|\boldsymbol{\theta} ,\textbf{y})\propto \frac{\pi(\textbf{x},\boldsymbol{\theta},\textbf{y})}{\tilde{\pi}_{GG}(\textbf{x}_{-i}|x_{i},\boldsymbol{\theta},\textbf{y})}\Big{|}_{\textbf{x}_{-i}=\textbf{x}_{-i}^{*}(x_{i},\boldsymbol{\theta})}\label{laplace}
\end{equation}
where $\textbf{x}_{-i}$ denotes the vector $\textbf{x}$ with the $i$th component omitted,  $\tilde{\pi}_{GG}(\textbf{x}_{-i}|x_{i},\boldsymbol{\theta},\textbf{y})$ is the Gaussian approximation of $\pi(\textbf{x}_{-i}|x_{i},\boldsymbol{\theta},\textbf{y})$ and $\textbf{x}_{-i}^{*}(x_{i},\boldsymbol{\theta})$ is the mode of  $\pi(\textbf{x}_{-i}|x_{i},\boldsymbol{\theta},\textbf{y})$. 
To obtained a simplified version of such a Laplace approximation $\tilde{\pi}_{SLA}(x_{i}|\boldsymbol{\theta},\textbf{y})$, which is defined as the series expansion  of $\tilde{\pi}_{LA}(x_{i}|\boldsymbol{\theta},\textbf{y})$ around $x_{i}=\mu_{i}(\boldsymbol{\theta})$, it is necessary to approximate the mode in the following way:
\begin{equation}\label{appro}
\textbf{x}_{-i}^{*}(x_{i},\boldsymbol{\theta})\approx \mathbb{E}_{\tilde{\pi}_{G}}(\textbf{x}_{-i}|x_{i})
\end{equation}
The conditional expectation (\ref{appro}) for Gaussian variables implies the following identity:
\begin{equation}
\frac{\mathbb{E}_{\tilde{\pi}_{G}}(x_{j}|x_{i})-\mu_{j}(\boldsymbol{\theta})}{\sigma_{j}(\boldsymbol{\theta})}=a_{ij}(\boldsymbol{\theta})\frac{x_{i}-\mu_{i}(\boldsymbol{\theta})}{\sigma_{i}(\boldsymbol{\theta})}
\end{equation}
for some $a_{ij}(\boldsymbol{\theta})$ when $j\ne i$. Denote $$x^{(s)}_{i}=\frac{x_{i}-\mu_{i}(\boldsymbol{\theta})}{\sigma_{i}(\boldsymbol{\theta})}.$$
Define then the following quantity, that we suppose exists:
$$d^{(3)}_{j}(x_{i},\boldsymbol{\theta})=\frac{\partial^{3} }{\partial x^{3}_{j}}\log{\pi(y_{j}|x_{j},\boldsymbol{\theta})}\Big{|}_{x_{j}=\mathbb{E}_{\tilde{\pi}_{G}}(x_{j}|x_{i})}.$$
\noindent The numerator and the denominator of expression (\ref{laplace}) can be expanded around $x_{i}=\mu_{i}(\boldsymbol{\theta})$ using the approximation (\ref{appro}) and the following lemma:
\begin{lem}
Let $\textbf{x}=(x_{1},\cdots, x_{n})^{T}\sim \mathcal{N}(\boldsymbol{0},\boldsymbol{\Sigma})$; then for all $x_{i}$
\begin{equation}
-\frac{1}{2}(x_{i},\mathbb{E}(\textbf{x}_{-i}|x_{i})^{T})\boldsymbol{\Sigma} ^{-1}(x_{i},\mathbb{E}(\textbf{x}_{-i}|x_{i}))=-\frac{1}{2}\frac{x^{2}_{i}}{\Sigma_{ii}}.
\end{equation}
\end{lem}
Approximating up to third order, we obtain for the $\log$ of the numerator of expression (\ref{laplace}):
\begin{multline}
\log{\pi(\textbf{x},\boldsymbol{\theta},\textbf{y})}\Big{|}_{\textbf{x}_{-i}=\mathbb{E}_{\tilde{\pi}_{G}}(\textbf{x}_{-i}|x_{i})}=-\frac{1}{2}\left(x_{i}^{(s)}\right) ^{2}+\\
+\frac{1}{6}\left( x_{i}^{(s)}\right) ^{3}\sum_{j\in \mathcal{I}\setminus{i}}d^{(3)}_{j}(\mu_{i}(\boldsymbol{\theta}),\boldsymbol{\theta})\left\lbrace \sigma_{j}(\boldsymbol{\theta}) a_{ij}(\boldsymbol{\theta})\right\rbrace ^{3}+\cdots.\label{eq:num}
\end{multline}
where $j \in \mathcal{I}\setminus{i}$ can take all values between 1 and $n$, except $i$. For the $\log$ of the denominator, we obtain:
\begin{multline}
\log{\tilde{\pi}_{GG}(\textbf{x}_{-i}|x_{i},\boldsymbol{\theta},\textbf{y})}\Big{|}_{\textbf{x}_{-i}=\mathbb{E}_{\tilde{\pi}_{G}}(\textbf{x}_{-i}|x_{i})}=\textrm{constant}-\\
-\frac{1}{2}x_{i}^{(s)}\sum_{j\in\mathcal{I}\setminus{i}}var_{\tilde{\pi}_{G}}(x_{j}|x_{i})d^{(3)}_{j}(\mu_{i}(\boldsymbol{\theta}),\boldsymbol{\theta}){\sigma_{j}(\boldsymbol{\theta})a_{ij}(\boldsymbol{\theta})}+\cdots ,\label{eq:gu}
\end{multline}
where
\begin{equation*}
var_{\tilde{\pi}_{G}}(x_{j}|x_{i})=\sigma^{2}\left\lbrace 1-corr_{\tilde{\pi}_{G}}(x_{i},x_{j})^{2}\right\rbrace .
\end{equation*}
Define the following quantities:
\begin{equation}
\gamma_{i}^{(1)}(\boldsymbol{\theta})=\frac{1}{2}\sum_{j\in\mathcal{I}\setminus{i}}var_{\tilde{\pi}_{G}}(x_{j}|x_{i})d^{(3)}_{j}(\mu_{i}(\boldsymbol{\theta}),\boldsymbol{\theta}){\sigma_{j}(\boldsymbol{\theta})a_{ij}}(\boldsymbol{\theta})\label{gam1}
\end{equation}
\begin{equation}
\gamma_{i}^{(3)}(\boldsymbol{\theta})=\sum_{j\in \mathcal{I}\setminus{i}}d^{(3)}_{j}(\mu_{i}(\boldsymbol{\theta}),\boldsymbol{\theta})\left\lbrace \sigma_{j}(\boldsymbol{\theta})a_{ij}(\boldsymbol{\theta})\right\rbrace ^{3}\label{gam3}
\end{equation}
then replacing (\ref{gam1}) in (\ref{eq:gu}) and (\ref{gam3}) in (\ref{eq:num}) we obtain
\begin{equation}
\log{\pi(\textbf{x},\boldsymbol{\theta},\textbf{y})}\Big{|}_{\textbf{x}_{-i}=E_{\tilde{\pi}_{G}}(\textbf{x}_{-i}|x_{i})}=-\frac{1}{2}\left(x_{i}^{(s)}\right) ^{2}+\frac{1}{6}\left( x_{i}^{(s)}\right) ^{3}\gamma_{i}^{(3)}(\boldsymbol{\theta})+\cdots
\end{equation}
and \begin{equation}
\log{\tilde{\pi}_{GG}(\textbf{x}_{-i}|x_{i},\boldsymbol{\theta},\textbf{y})}\Big{|}_{\textbf{x}_{-i}=E_{\tilde{\pi}_{G}}(\textbf{x}_{-i}|x_{i})}=\textrm{constant}-x_{i}^{(s)}\gamma_{i}^{(1)}(\boldsymbol{\theta})+\cdots
\end{equation}
Finally we have:
\begin{eqnarray}
&&\log{\tilde{\pi}_{SLA}(x_{i}^{(s)}|\boldsymbol{\theta},\textbf{y})}=\log{\pi(\textbf{x},\boldsymbol{\theta},\textbf{y})}-\log{\tilde{\pi}_{GG}(\textbf{x}_{-i}|x_{i},\boldsymbol{\theta},\textbf{y})}= \\ \nonumber
&=&\textrm{constant}-\frac{1}{2}\left( x_{i}^{(s)}\right) ^{2}+\gamma_{i}^{(1)}(\boldsymbol{\theta})x_{i}^{(s)}+\frac{1}{6}\left( x_{i}^{(s)}\right) ^{3}\gamma_{i}^{(3)}(\boldsymbol{\theta})+\cdots.
\end{eqnarray}
\item[Third step] This step combines the previous two steps with numerical integration:
\begin{equation}
\tilde{\pi}(x_{i}|\textbf{y})=\sum_{k}\tilde{\pi}_{SLA}(x_{i}|\theta_{k},\textbf{y})\tilde{\pi}{(\theta_{k}|\textbf{y})}\Delta_{k}.\label{appr}
\end{equation}
The sum is over values of $\boldsymbol{\theta}$ defined in the first step with area weights $\Delta_{k}$.
\end{description}
Besides the explicative model (\ref{eq:eta}) we also include a descriptive model in the analysis which  takes into account only
spatial autocorrelation: $\textrm{logit}(\pi_{i})=\mu + f^{(s)}_{d_{i}}$ or assume  "convolution prior" for the
spatial components: $\textrm{logit}(\pi_{i})= \mu + f^{(s)}_{d_{i}} + f^{(u)}_{d_{i}}$. In the best of the two
previous models, we add in several explicative models different potential effects, in $\textbf{f}^{(a)}$: distance or access time
to the healthcare provider, distance between geographical unit of residence and the second healthcare provider and medical density of the geographical unit. The proximity zones are added using indicator variables (the zone where the healthcare provider under study is, is used as the reference zone): $\sum^6{\beta_{k}\mathbb{I}(i \in Z_k)}$ where $\mathbb{I}(i \in Z_k)=1$ only if the geographical unit $i$ belongs to the proximity zone $Z_k$. The \texttt{INLA} \texttt{R} package is used for implementing models. The goodness
of fit of each model is assessed using the deviance information criterion (DIC) \citep{Spiegelhalter2002}, as a generalization of
the Aka\"{i}ke score. We also use DIC for comparing models. The DIC is defined as $DIC=\bar{D}+p_D$, decomposed like
penalized likelihood indicators into two terms: $\bar{D}$ measuring the fit to data and $p_D$ measuring the
complexity of the models.

\section{Results}
Each model needs less than 30 seconds in \texttt{R}, on a PC with a 2.29 GHz dual core processor (compared to 2 or 3 hours using fully Bayesian inference).

\subsection{Descriptive models}
\begin{figure}[H]
 \centering
 \includegraphics[scale=0.7]{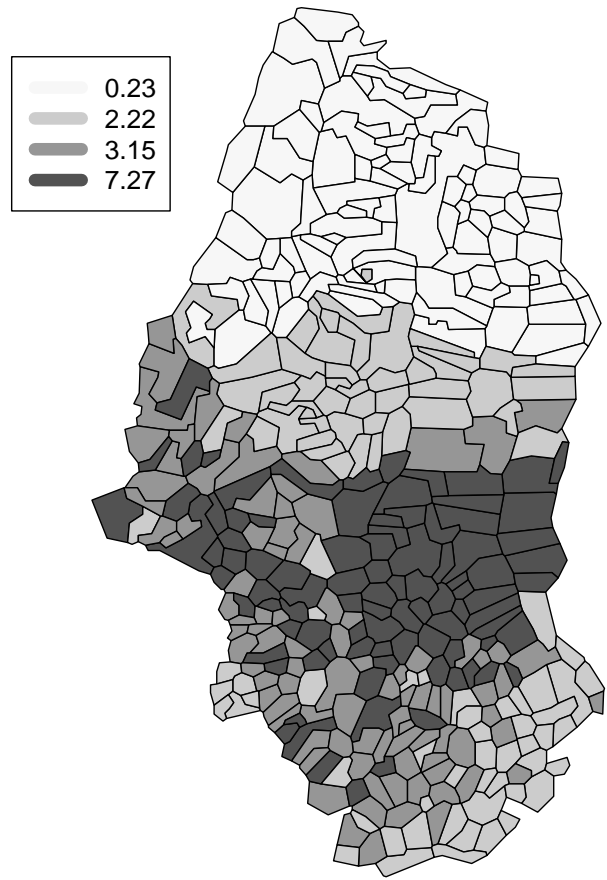}
 \caption{\textbf{Exponential of the spatial effect (ICAR), descriptive model}}
 \label{fig:ICAR}
\end{figure}

We compare the ICAR-only model with the convolution prior model. The DIC of the former is 2253.6 (for an effective number of parameters of 230.6) and the DIC of the later is 2253.5 (and 231.5), indicating that the first model is good enough on the dataset. Figure (\ref{fig:ICAR}) is the exponential of the ICAR spatial effect (readable as a relative risk).

\subsection{Explicative models}

\begin{table}[H]
\caption{\label{tab:DIC}DIC of the different explicative models}
\centering
\fbox{%
\begin{tabular}{lrr}
 \multicolumn{1}{c}{Model} & \multicolumn{1}{c}{$p_D$} & \multicolumn{1}{c}{DIC}\\
\hline
  ICAR alone & 230.6 & 2253.6 \\
  ICAR and distance to provider & 224.7 & 2254.3\\
  ICAR and access time to provider & 224.6 & 2245.1\\
  ICAR and distance to the second provider & 203.5 & 2241.1\\
  ICAR and proximity zone (as factor) & 196.7 & 2234.5\\
  ICAR and medical density & 231.6 & 2254.6\\
\end{tabular}}
\end{table}

\begin{figure}[H]
 \centering
 \includegraphics[scale=0.6]{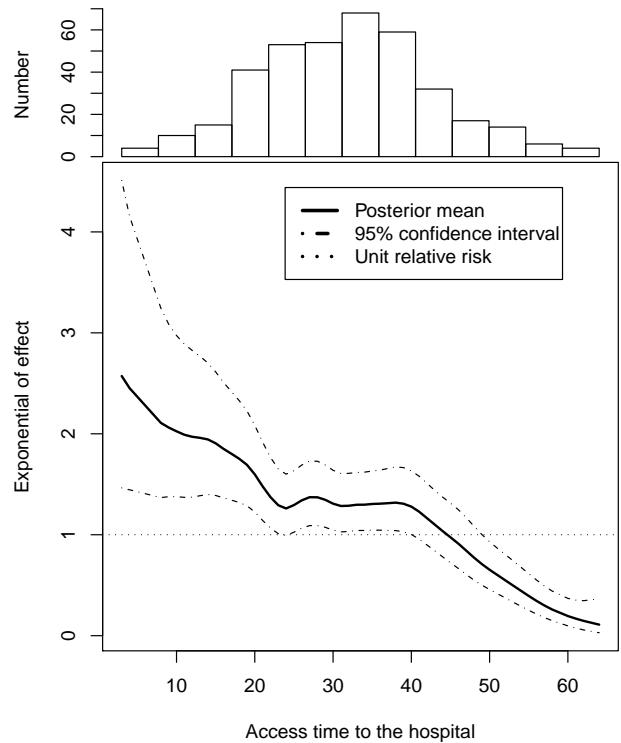}
 \caption{\textbf{Exponential of the access time effect, explicative model}}
 \label{fig:temps}
\end{figure}

Table (\ref{tab:DIC}) summarizes the DICs of different models. The DIC indicates that in addition to spatial effect by an ICAR prior, taking into account the distance to provider (surprisingly) or practitioner density, does not improve the model. On the contrary, in addition to the ICAR prior, taking into account the access time between the geographical unit of patient residence and the healthcare provider improves the model compared with the similar model with only the ICAR prior (DIC increases by 7.5). In Figure (\ref{fig:temps}), the relative risk decreases for time less than 20 minutes but remains greater than 1, then decreases again to a value below 1 after 40 minutes.

\begin{figure}[H]
 \centering
 \includegraphics[scale=0.6]{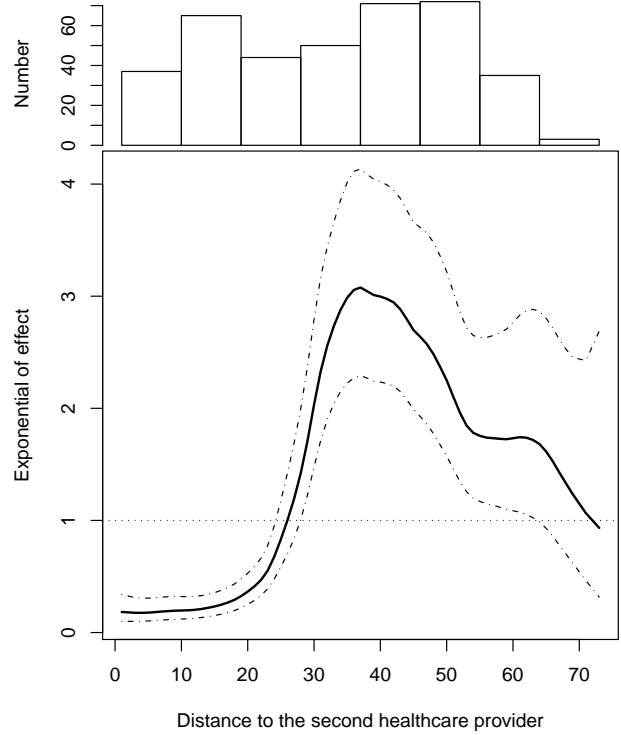}
 \caption{\textbf{Exponential of the distance to second healthcare provider effect, explicative model, (same legend as Figure (\ref{fig:temps}))}}
 \label{fig:dHCC}
\end{figure}

The distance to the second provider also improves the initial model and Figure (\ref{fig:dHCC}) shows the major concurrent effect of this second provider on the first one when patients live less than 20 kilometers away from it.

\begin{figure}[H]
 \centering
 \includegraphics[scale=0.7]{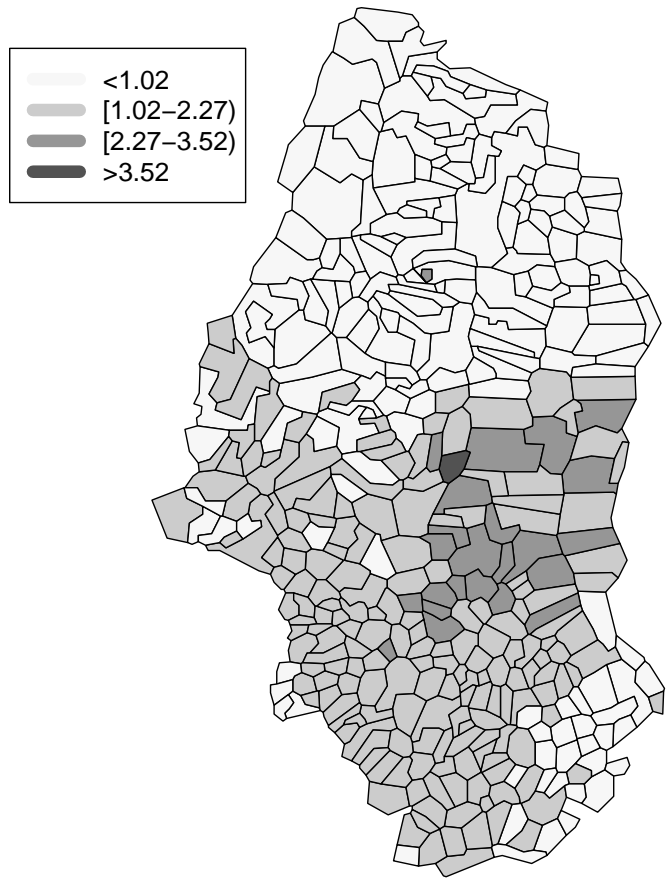}
 \caption{\textbf{Exponential of the proximity zone adjusted spatial effect, explicative model}}
 \label{fig:Zp}
\end{figure}

Table (\ref{tab:DIC}) shows, furthermore, that the proximity zone greatly improves the initial model. We use fixed effect and random effect for this covariate but as the results are slightly similar, only the results with fixed effect are shown. Table (\ref{tab:Zp}) summarizes these results (with respect to the zone number 7 where the healthcare provider is situated). Figure (\ref{fig:Zp}) shows that, even adjusted to zone, an over-recruitment persists in the zone number 7 and an under-recruitment in the north (role of the second healthcare provider) but also in a small South-East sub-region.

\begin{table}[H]
\caption{\label{tab:Zp}Proximity zone (fixed effect)}
\centering
\fbox{%
\begin{tabular}{lccc}
  & Posterior mean & \multicolumn{2}{c}{95\% confidence interval} \\
\hline
  1 Altkirch & 0.83 & 0.63 & 1.07\\
  2 Colmar & 0.08 & 0.05 & 0.13\\
  3 Guebwiller & 0.39 & 0.29 & 0.52\\
  4 Saint-Louis & 1.07 & 0.79 & 1.44\\
  5 Sélestat & 0.04 & 0.01 & 0.11\\
  6 Thann & 1.24 & 0.92 & 1.66\\
  7 Mulhouse & \multicolumn{3}{c}{Reference zone}\\
\end{tabular}}
\end{table}

\section{Discussion}

Introduction of covariates in the models we tested is straightforward but we must keep in mind that nothing is gained with the INLA technique if the number of these covariates is more than 6. If this is the case, the McMC technique remains the most useful technique. Furthermore, INLA relies on a latent Gaussian model and in order to smooth the effect of distance, we used a random walk. A powerful alternative of random walks is to use splines~\citep{Ruppert2003}. Among different splines, linear combinations of B-splines~\citep{Eilers1996} have useful properties and offer a lot of flexibility but even a certain "wiggliness". A solution is to penalize second derivatives or differences on the coefficients of the linear combination~\citep{Eilers1996}. We then obtain P-splines. In the Bayesian framework, the stochastic equivalent of differences are random walks. For example, following~\citet{Lang2004}, if $B$ are B-splines of order $d$, a P-spline is defined by $\sum_{m=1}^{d+k}{\beta_m B_m}$, assuming $k$ regularly spaced knots. Priors on $\beta$ are then random walks of first or second order with gaussian errors $\epsilon_m \sim{N(0, \frac{1}{\tau_{\epsilon}})}$. Priors on $\beta_1$ and eventually $\beta_2$ are flat (uniform distributions) and gamma prior is assumed on the precision $\tau$. But the estimation of this kind of model using INLA is not possible. P-splines can be reformulated as a latent Gaussian model (see for example~\citet{Crainiceanu2005}) but there are as many parameters to be estimated as the number of spline knots. The fully Bayesian framework using McMC is hence the better approach, using for example \texttt{BayesX} specifically devoted to StAR models or, with more difficulty, \texttt{BUGS}. In our case using random walks as prior on effects was not an important limitation as it was easy and rational to round or categorize our variables. Of course, outside the latent Gaussian model framework needed for INLA and in addition to the ICAR prior for spatial effect, relying on adjacency between geographical units, several other model can be used for spatial smoothing~\citep{Kammann2003}, for example bidimensional P-splines~\citep{Lang2004} (or more generally~\citet{Wood2006}) can be fitted on the centroids of the geographical units, as implemented in \texttt{BayesX}.

We consider here that the number of people at risk is the population of a geographical unit. But we could also apply to this population a factor representing the proportion of the population that can be recruited. This "hospitalizability" is different according to the concerned pathology: e.g. 20\% of the total population or 30\% of men over 75 years, based on the prevalence of diabetes. Rather than consider expected number as a fixed percentage of the population, it is logical to try to adjust this percentage, for example, by age. This echoes the traditional techniques of
standardization of risk used, for example, in disease mapping.

Eventually \citet{Rue2009} described two useful methods for approximating $\pi(x_{i}|\boldsymbol{\theta},\textbf{y})$ in the equation (\ref{Bayes}). In this paper we describe and use a simplified Laplace approximation but two Laplace approximations in the equation (\ref{laplace}) can be used instead of. The accuracy of the simplified Laplace approximation can be not good enough for the computation of predictive measures (like conditional predictive ordinate or cross-validated probability integral transform) and the full Laplace approximation has sometimes to be used \citep{Held2010}.

An alternative way, besides splines or random walks for modeling the attenuation of the recruitment with distance $d$, can be to use a generalization of the Reilly distribution. Recruitment can then vary with $\frac{1}{d^{\rho}}$, where $\rho$ is a parameter to be estimated. In the case of binomial distribution for the response variable, we need to transform $\frac{1}{d^{\rho}}$ on the logit scale and hence the function to be included in the models is $-\ln(d^\rho-1)$. In a Bayesian framework we would assume a vague prior distribution on $\rho$, for example a uniform on 0 to 5. We tested the use of this generalized-Reilly distribution in our application (unpublished manuscript) but it seems that these models are not flexible enough compared with smoothing by random walks. On the other hand, Reilly is parametric function and an estimation of the function parameter can be retrieved from the model and easily interpreted.
An important limitation of the approach using aggregated data and Poisson or binomial models is that an observation in the dataset is a set of covariates related to the geographical unit and potentially also to some other variables which need to be categorized. For example, in our application if we are interested in the recruitment taking into account the age of patients, we have to categorize the age and count the number of patients in all the combinations of geographical units and age categories. A powerful approach is then to use, for example, Poisson-kriging models \citep{Goovaerts2006,Goovaerts2008}.
The attenuation of the recruitment according to the distance or access time, can be isotropic (the recruitment is hence the same on all points of a circle around the healthcare provider) but can also be anisotropic. For example, we can assume that the distance effect will not be the same in all of the proximity zones, $Z$, and build a varying coefficient model including some terms $\mathbb{I}(i \in Z_k) \cdot f(d_i)$, where $f(d_i)$ can be modeled using random walks.

\section{Acknowledgments:} The third author was supported by the project {\em start up giovani ricercatori} of the University of Cagliari (Italy). The authors thank Frank McKenna for his careful review of the manuscript.

\bibliography{INLA_SRR}
\bibliographystyle{chicago}
\end{document}